\documentclass{cargese}
\usepackage{graphicx}
\usepackage{amsmath}
\usepackage{amssymb}
\let\footnote\savefootnote
\let\footnotetext\savefootnotetext
\let\footnoterule\savefootnoterule
\normallatexbib
\def\om{\omega}
\def\bk{\mathbf{k}}

\begin{document}
%------------ article title  ------------------->>
% For a long title use \\ to cut lines.
% In that case, supply  alternate version of the title
% in square brackets, (it will go in the Table of contents during final
% production of the book.
% \articletitle[Short title]{The long version \\ of this title}

 \articletitle[Black Hole Production] {Topics in Black
Hole Production}

%% optional, to supply a shorter version of the title for the running head:
%%\chaptitlerunninghead{}

\author{Vyacheslav S.~Rychkov\protect{\footnote{To appear in the proceedings of the Carg\`ese Summer School
``String Theory: From Gauge Interactions to Cosmology", June 7-19,
2004.}}}

%% multiple authors at the same institution may be separated with \\
%% like in \author{Samuel Bostaph\\
%%                 and Gregor Kariotis}

%% Your Institution and address. May cut into  separated lines with \\

\affil{Institute for Theoretical Physics, University of Amsterdam,\\
1018XE Amsterdam, The Netherlands}

% optional email address
%\email{}

%% Repeat the above for multiple authors at different institutions.
%% \author{ }
%% \affil{ }
%% \email{ }

% optional abstract
\begin{abstract} We revisit Voloshin's model of multiple black
hole production in trans-Planckian elementary particle collisions
in $D=4$. Our revised computation shows that the cross section to
produce $n$ additional black holes is suppressed by $s^{-1}$,
rather than being enhanced as was originally found. We also review
the semiclassical gravity picture of black hole production from
hep-th/0409131, making additional comments about the meaning of
wavepacket subdivision.
\end{abstract}

%------------ body of article ------------------->>
% Write your article here.
% Note that the \section{section title}
% command allows for the form \section[short title]{very long\\ title}
% Idem for \subsection and \subsubsection
%------------ end of article ------------------->>

\section{Introduction}

Black hole (BH) production in trans-Planckian elementary particle
collisions ($E\gg E_{Planck}$) has long been considered a
theoretical possibility. %\cite{tHooft}.
If TeV-scale gravity
scenarios based on large extra dimensions %\cite{ADD}
or warped compactifications
%\cite{RS}
are realized in nature, this possibility may be realized in
practice at future accelerators (see \cite{Kanti} for a recent
review).
%\cite{GT}
The key question is the cross section of this
process, which is usually assumed to be set by the horizon radius
of the produced BH (the so called ``geometric cross section"
$\sigma\sim \pi r_h^2$).
%\cite{BF}).

In this note I would like to discuss two aspects of the BH
production problem relevant for justifying the geometric cross
section estimate. First I will explain how to derive this estimate
from a controlled semiclassical gravity approximation to the BH
production process, adding some comments to the original
discussion of \cite{GR} (also recently reviewed in \cite{R3}).
Then I will estimate the cross section of multiple BH production
due to collisions between virtual gravitons emitted by the primary
particles, in a model first proposed by Voloshin \cite{Vol}. The
conclusion
%\cite{RV}
(contrary to \cite{Vol}) is
that multiple BH production gives a subdominant contribution to
the total cross section.

%A good review of BH production phenomenology in large extra
%dimension scenarios is \cite{Kanti}.
The size of produced BHs in large extra dimension scenarios is
typically much smaller than the size of extra dimensions, and
their production may thus be considered as happening in flat $D$
dimensional spacetime. Since we are focussing on theoretical
issues, to keep the discussion clear we will work in $D=4$. We
will also use Planck units, setting $E_{Planck}=1$.

\section{Wavepackets and Semiclassics}

In this note we will adhere to the standard believe that in
quantum theory of gravity classical BHs with mass $M\gg 1$ will be
realized as long-lived resonance states, decaying via Hawking
radiation, with lifetime $\sim M^3$. It is then energetically
allowed to produce such BHs in trans-Planckian ($E\gg1$)
elementary particle collisions.

To estimate the cross section of this process, Eardley and
Giddings \cite{EG} looked at the grazing collision of two
ultrarelativistic point particles in classical general relativity,
using formation of a closed trapped surface (CTS) as a sufficient
condition for BH formation. In this totally classical description,
lower bound for the cross section is given by $\pi b_{max}^2$,
where $b_{max}$ is the maximal impact parameter for which we are
able to find a CTS in the spacetime formed by two colliding
Aichelburg-Sexl shock waves. \cite{EG} found $b_{max}\sim E\sim
r_h$, and this implies $\sigma\sim \pi r_h^2$. Recourse to such an
indirect method is necessary, because explicit solutions of
Einstein's equations exhibiting the final BH state are out of
reach.
%(Work is now in progress to obtain the most optimal
%estimates achievable by this method, by pushing the CTS to the
%boundary of the known part of the collision spacetime \cite{YR}.)

 How do we justify this approach from quantum gravity point of view? A point of immediate concern is that
particles in a collider experiment are described by wide
wavepacket states of macroscopic size (set essentially by the beam
radius, which in turn is determined by the focussing ability of
accelerator magnets). These wavepackets are vastly larger than BHs
whose production we are trying to describe. If we use the energy
momentum tensor of these wavepacket states $|\psi \rangle$ in the
RHS of the semiclassical Einstein's equations
\begin{align}\label{semi}
    R_{\mu\nu}-\textstyle\frac 12 g_{\mu\nu} R = \langle \psi| T_{\mu\nu}|\psi
    \rangle,
\end{align}
we won't see any BH production whatsoever, since the energy is
spread out over a huge volume, and the energy density is
insufficient to cause collapse. Informally speaking, particles
``do not fit" inside a BH.

However, this does not mean that BHs do not form. The correct
interpretation is that the part of the gravitational field
wavefunction corresponding to BH production was erased---averaged
away---by eq.~(\ref{semi}). Some averaging is always inevitable
when using the semiclassical field equations, since we substitute
$T_{\mu\nu}$ by its expectation value. Unfortunately, in this case
it destroys precisely the part of the wavefunction we are
interested in.

\begin{figure}
\begin{center}
\includegraphics[height=2cm]{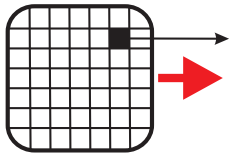}
 \caption{Subdivision of the right-moving particle
wavepacket. Similar subdivision has to be done for the left-moving
particle.} \label{subdiv}
\end{center}
\end{figure}

To see BH production, one should instead proceed as follows. First
of all, we have to subdivide the initial wavepackets into much
smaller wavepackets of size $w\ll r_h$:
\begin{align}\label{}
    |\psi\rangle = N^{-1/2}\sum\nolimits_{i=1}^N|\psi_i\rangle.
\end{align}
This subdivision is carried out so that the small wavepackets
$|\psi_i\rangle$ in the RHS are almost orthogonal. This
orthogonality is quite obvious in the position representation (see
Fig.~\ref{subdiv}). Because of the orthogonality, collisions
between different pairs of small wavepackets are mutually excluded
possibilities, and probabilities of BH production in each such
elementary collision should be added. Now, it is the collisions of
the small wavepackets that we are going to analyze using
eq.~(\ref{semi}). Condition $w\ll r_h$ ensures that the small
wavepackets produce a collision spacetime which is a small
perturbation of the one corresponding to point particles of the
same energy. Thus the Eardley-Giddings analysis applies, and
adding probabilities results in the geometric cross section.

Using finite-size wavepackets instead of point particles has an
additional bonus in that it puts the conditions of applicability
of the semiclassical approximation under control. For example,
curvature blows up when the shock fronts of Aichelburg-Sexl waves
corresponding to point particles collide \cite{R1,R2}. However,
taking wavepacket size into account regulates the curvature and
brings it below the Planck value, so that we can trust the
semiclassical gravity approximation \cite{GR}.

\section{Multiple Black Hole Production}

According to the above discussion, the geometric cross section
formula provides a lower bound for a single large BH production
cross section in a trans-Planckian collision. However, as the
energy of the particles grows, multiple BH production also becomes
energetically allowed. It is important to understand which process
is dominant at asymptotically high energies. If multiple BHs
dominate, it will be much harder to observe dipole patterns of
emitted particles expected in the Hawking evaporation of a single
large BH.

\begin{figure}
\begin{center}
\includegraphics[height=2cm]{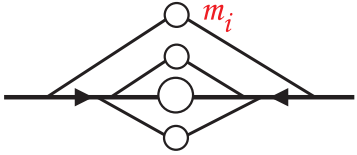}
 \caption{A typical diagram for multiple BH production in Voloshin's model} \label{multiple}
\end{center}
\end{figure}

We are going to discuss a model proposed by Voloshin \cite{Vol},
in which multiple BHs are produced due to collisions between
virtual gravitons emitted by the trans-Planckian projectiles. Such
virtual graviton emission is a quantum effect: classically, any
radiation happens after the particles collide. The process is
studied diagrammatically, with a typical diagram shown in
Fig.~\ref{multiple}. Only the case of small peripheral BHs ($1\ll
m_i\ll E$, $i=1\ldots n$) is considered, so that the gravitons are
``soft".

\begin{figure}[b]
\begin{center}
\includegraphics[height=1cm]{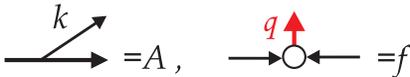}
 \caption{Graviton emission and BH production vertices} \label{vertices}
\end{center}
\end{figure}

 The production amplitude is computed from the diagrams using the
standard QFT propagators and particular vertices for soft graviton
emission and for BH production (Fig.~\ref{vertices}). The
amplitude to emit a positive helicity graviton of energy $\om\ll
E$ and small transverse momentum $\bk=(k_2,k_3)$, $|\bk|\ll\om$ is
given by (see Appendix A)
\begin{align}\label{emission}
    A \propto (E/\om)^2(k_2+ik_3)^2\ .
\end{align}
(We will not pay attention to constant numerical factors. Thus our
final result (\ref{ampl}) is valid up to a factor $c^n$.)

Although we are unable to compute the elementary BH production
vertex $f(q^2)$, the geometric cross section allows to fix the
combination
\begin{align}\label{comb}
    |f(q^{2})|^2 \rho(q^2) \sim (q^2)^2,
\end{align}
where $\rho(q^2)$ is the density of BH states at mass
$\sqrt{q^2}$.

A crucial final element of the model is a condition which ensures
that the emitted gravitons do not subsequently fall into a common
large BH. Such an infall may happen due to graviton rescattering
diagrams, which we are not going to compute. Thus, without a
``fall safe" condition we would be in danger of greatly
overestimating the multiple BH amplitude. Voloshin's ``fall safe"
condition limits the transverse momenta of emitted gravitons by
$|\bk|\lesssim 1/E$.

To derive this condition, we note that a typical emitted graviton
will be off-shell by $\Delta E \sim \bk^2/\om$. It will exist for
a time interval $\Delta t\sim 1/\Delta E$, during which it will
reach transverse separation $\Delta z\sim (\bk/\omega)\Delta t $
from the projectile. Voloshin's condition arises if we require
that this transverse separation is larger than the horizon radius
of the big BH formed in the collision of the primary particles:
$\Delta z \gtrsim E$.

In the described model, our computation (see Appendix B) gives the
following amplitude to produce $n$ additional BHs with 4-momenta
$q_i$:
\begin{align}\label{ampl}
    f^{(n)}(s,q^2_i)\sim f(s)\prod\nolimits_{i=1}^n
    (q_i^2)^{-2}f(q_i^2)\ .
\end{align}
The original computation of \cite{Vol}, Eq.~(5), gave an amplitude
larger than (\ref{ampl}) by a factor of $(s\hspace{1pt} q_i^2)^n$.
We believe that our result is correct; see Appendix B and
\cite{Erratum} for an explanation.

 Using (\ref{comb}), we can compute from (\ref{ampl}) the contribution of the diagrams from Fig.~\ref{multiple}
 into the total cross section $\sigma_n$ to produce
 one large ($m^2\sim s$) and $n$ small BHs. This contribution will behave like $s^{1-n}$, the suppression
 being due to the phase space
 restriction $|\bk|\lesssim 1/E$ satisfied by the small BHs as a consequence of the ``fall safe" condition.

\begin{figure}[t]
\begin{center}
\includegraphics[height=1.5cm]{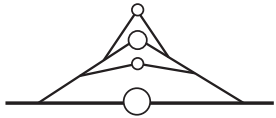}
 \caption{A diagram giving an $s$-independent contribution to $\sigma_n$} \label{leading}
\end{center}
\end{figure}

However, for $n\ge 2$ there are diagrams which give a larger
contribution, so that $\sigma_n\sim const$ is likely for any $n\ge
1$. Consider, e.g., Fig.~\ref{leading},
 where the primary particles emit ``fall safe" gravitons of energy $E_1$, $1\ll E_1\ll E$, and it is these gravitons that form $n$
 smaller BHs according to the previous model. The allowed phase space for this diagram will be much bigger,
 since the individual small BHs can now have much larger transverse momenta $|\bk|\lesssim 1/E_1$, only
 their sum being $\lesssim 1/E$. Choosing $E_1$ above the threshold of $n$ BH production,
 we will get an $s$-independent contribution to $\sigma_n$.\footnote{It is also easy to see that the \emph{inclusive} cross section of multiple BH production cannot decay with $s$.
 This is because the particles may first reduce their energy by emitting one or more ``fall safe" gravitons
 (which costs no $s$-dependent factor, see \cite{Vol}, Eq.~(3)), and then collide to form BHs.  I am grateful
 to M.~Voloshin for this remark.}

In any case, we see that $\sigma_n$ is suppressed compared to the geometric cross section value $\sigma_0\sim s$.
\section{Conclusions}

In this note we discussed two aspects of the BH production
problem. In Section 2 we discussed how wavepacket arguments can be
used to justify the use of semiclassical gravity in this problem.
In Section 3, we revisited Voloshin's model of multiple BH
production, and presented a revised computation which shows that
this process is suppressed compared to the production of a single
large BH. Our main conclusion is that the geometric cross section
estimate seems to be in rather good health, surviving all checks
and resisting any disproving attempts.

%% optional
\begin{acknowledgments}
I would like to thank Steve Giddings for the opportunity to
collaborate on \cite{GR}. A part of Section 2 arose as an answer
to Marco Cavagli\`a's interesting questions. I am especially
grateful to Misha Voloshin for the e-mail correspondence which
helped me to understand his model, and for the critical remarks
about the early versions of the computation. It is also a pleasure
to thank the organizers of the Carg\`ese 2004 String Theory
school. This work was supported by Stichting FOM.

\end{acknowledgments}

\chapappendix{A. Soft graviton emission vertex} Consider emission
of a positive helicity collinear graviton with energy $\omega$ and
small transverse momentum $\bk=(k_2,0)$, $k_2\ll \omega$. Its
polarization tensor, satisfying the constraints $h_{ij}k_j=0$,
$h_{ii}=0$, is given by
\begin{equation}\label{}
    h_{ij}\approx\begin{pmatrix}
 k_2^2/\omega^2 & -k_2/\omega &  -ik_2/\omega \\
    -k_2/\omega& 1 &  i  \\
    -ik_2/\omega & i & -1-k_2^2/\omega^2  \\
    \end{pmatrix}\ .
\end{equation}
The emission amplitude is $A\propto \langle p-k|T_{ij}|p\rangle
h_{ij}$, where the energy momentum tensor matrix element is
\begin{equation}\label{}
    \langle P|T_{\mu\nu}|p\rangle=\langle P|\phi_{,\mu}\phi_{,\nu}-\frac 12 \eta_{\mu\nu}(\partial\phi)^2|p \rangle=p_\mu
    P_\nu+p_\nu P_\mu-\eta_{\mu\nu}(pP)\,.
\end{equation}
From this we find $A\propto{E^2k_2^2}/{\om^2}$.

\begin{figure}[h]
\begin{center}
\includegraphics[scale=0.3]{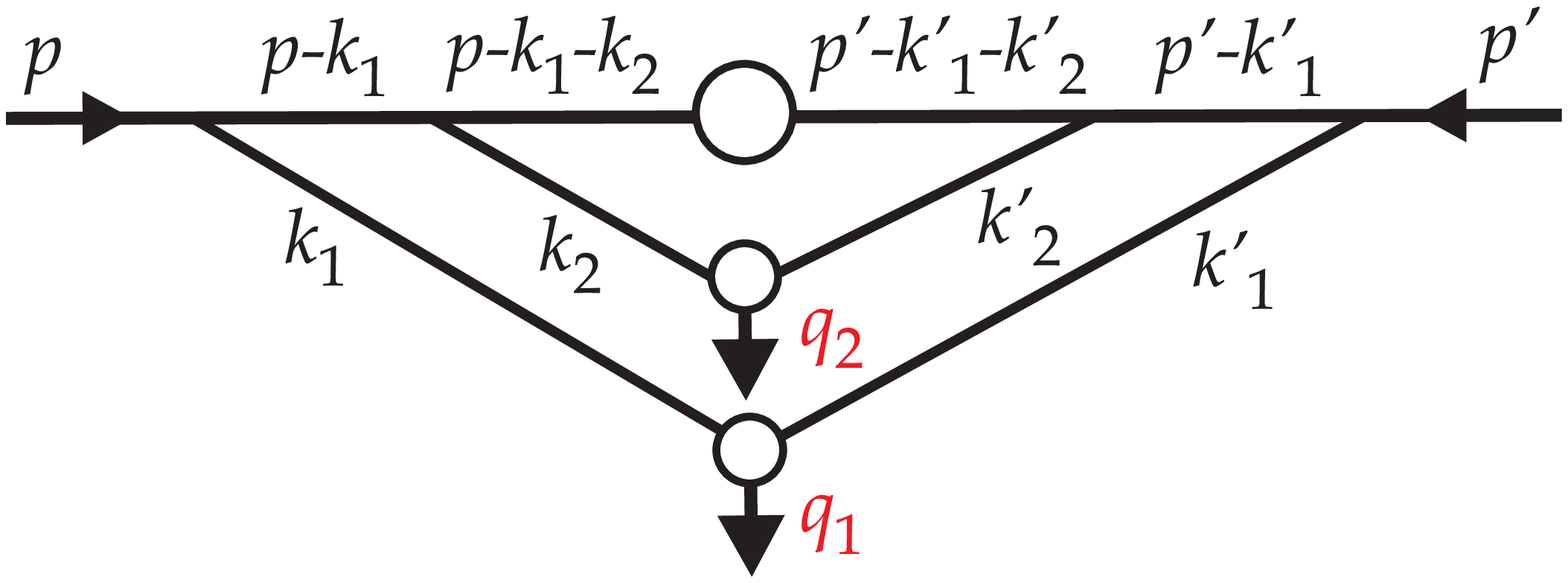}
 \caption{A diagram for one large and two small BHs in the final state} \label{2BH}
\end{center}
\end{figure}

\chapappendix{B. Multiple BH production amplitude} We will compute
the $n=2$ amplitude in the case when both small BHs are produced
at rest in the c.m.~frame: $q_i=(m_i,0,0,0)$ in Fig.~\ref{2BH}.
The calculation of the amplitude naturally splits into 3 steps.
First, we compute the loop integral over the longitudinal momenta.
Then, we multiply by the emission vertices and integrate over the
transverse momenta. Finally, we multiply by the BH production
vertices.

It is convenient to write the graviton momenta as follows
($i=1,2$):
\begin{align}\label{7}
    k_i =(\frac {m_i} 2+\frac{x_i-y_i}2, \frac {m_i} 2-\frac {x_i+y_i}2,\bk_i),\quad
    k_i'=q_i-k_i\ .
\end{align}
%In this parametrization the graviton and projectile virtualities
%are:
%\begin{align}
%    k_i^2&\approx m x_i-\bk_i^2,&  k_i'{}^2&\approx
%    my_i-\bk_i^2\ ,\notag\\
%    (p-k_1)^2&\approx -E x_1-\bk_1^2,& (p'-k_1')^2&\approx
%    -Ey_1-\bk_1^2\ ,\label{virt}\\
%    (p-k_{12})^2&\approx -E x_{12}-\bk_{12}^2,& (p'-k_{12}')^2&\approx
%    -Ey_{12}-\bk_{12}^2\ ,\notag
%\end{align}
%where $k_{12}=k_1+k_2$ etc. A Since we ultimately impose
%Voloshin's fall safe condition $\bk \ll 1/E$, we see that omission
%of terms quadratic in $x,y$ from (\ref{virt}) is justified. For
%the same reason
%
The longitudinal loop integral now separates into the integrals
over $x_i$ and $y_i$. The part depending on $x_i$ is
\begin{align}
    I=&\int dx_1\, dx_2 \frac 1{k_1^2 k_2^2
    (p-k_1)^2 (p-k_1-k_2)^2} \label{int}\\
    \approx&\int \frac {dx_1\, dx_2}{(m_1 x_1-\bk_1^2)(m_2 x_2-\bk_2^2)(-E x_1-\bk_1^2)(-E
    (x_1+x_2)-\bk_{12}^2)}\ .\notag
\end{align}
The $+i\epsilon$ is implicit in each denominator. This integral is
easy to compute by closing the contour in the lower half-plane.
Omitting corrections of the order $m/E\ll 1$, we have
\begin{align}\label{}
    I\propto [E^2 \bk_1^2(\bk_1^2+\bk_2^2)]^{-1}\ .
\end{align}

 Since we used complex analysis, it is important to check that the
integral is dominated by soft, almost real gravitons.  This is
indeed true, the important region being $|x_i|\lesssim
\bk_i^2/m_i$. The tail corresponding to $|x_i|\gg \bk_i^2/m_i$ can
be estimated directly as:
\begin{align}\label{}
    \frac 1{m_1 m_2 E^2}\int \frac{dx_1}{x_1^2} \frac{dx_2}{x_2^2}\ll
    I\ .
%m^{-2} E^{-2}\textstyle\int (dx_1/x_1^2)\,(dx_2/x_2^2)\ll I\ .
\end{align}
This check also justifies \emph{post factum} neglecting the
dependence of the graviton emission vertices on $x_i$, as well as
the omission of $x^2$ terms in the denominators of (\ref{int}).

\begin{figure}[h]
\begin{center}
\includegraphics[scale=0.2]{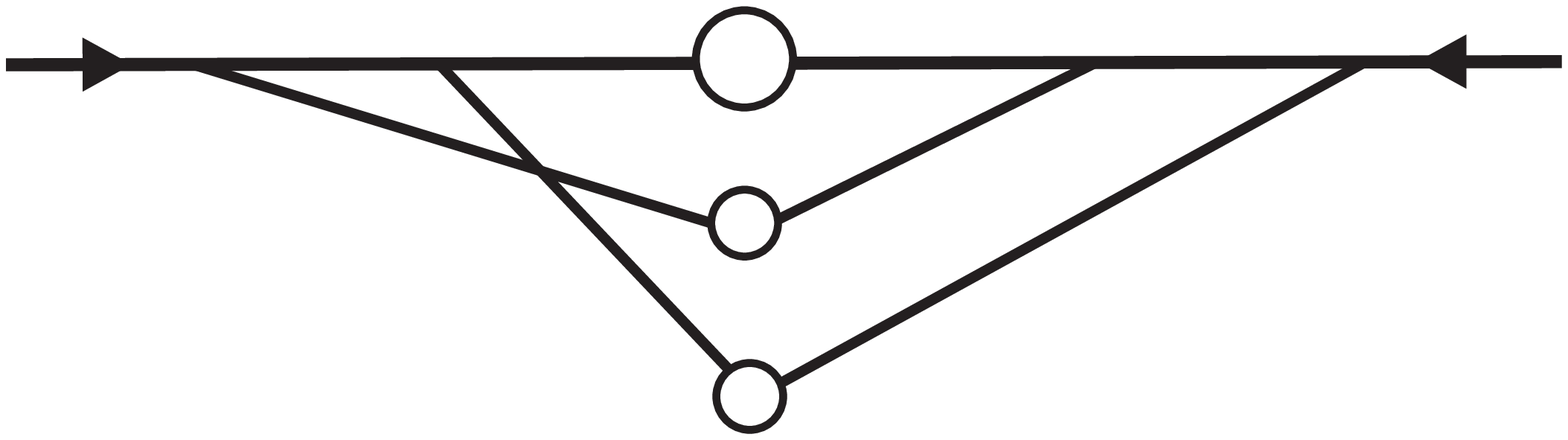}
 \caption{One of the 3 permuted diagrams} \label{perm}
\end{center}
\end{figure}

Before proceeding to the next step, we have to add diagrams
differing by the order of graviton emission (Fig.~\ref{perm}).
This summation has the effect $I\to [E^2 \bk_1^2\bk_2^2]^{-1}$.
Finally, we multiply by the same factor arising from the
$y$-integration, and get:
\begin{align}\label{long}
[E^2\bk_1^2\bk_2^2]^{-2}\quad \mbox{(longitudinal part)}\ .
\end{align}
%% appendix optional
%\chapappendix{This is the Appendix Title}
%This is an appendix with a title.
%\chapappendix{}
%This is an appendix without a title.

We will assume that the small BHs are produced in the spin 0
state, so that the colliding gravitons have opposite helicities.
The product of the corresponding emission vertices is
\begin{align}\label{}
\propto (E/m)^4(k_2+ik_3)^2(k_2-ik_3)^2= (E/m)^4(\bk^2)^2\ .
\end{align}
Multiplying (\ref{long}) by such factors for both BHs and
integrating over $|\bk_i|\lesssim 1/E$ (the ``fall safe"
condition), we get a number $\sim (m_1 m_2)^{-4}$.

 Finally, we multiply by the BH production vertices and arrive at
\begin{align}\label{}
    f^{(2)}\sim f(s)\,f(m_1^2)\, f(m_2^2)\, (m_1 m_2)^{-4}\ .
\end{align}
This formula agrees with the general result (\ref{ampl}) in the
considered case.

Extension of the above computation to the general case is quite
straightforward. A point worth mentioning is the use of the
standard identity
\begin{align}\label{}
    \sum\nolimits_{perm}[a_1(a_1+a_2)\cdots
    (a_1+\ldots+a_n)]^{-1}=[a_1\cdots a_n]^{-1}
\end{align}
when summing over the order of graviton emission in the $n>2$
case.

In \cite{Vol}, the estimate $\int d^4 k_i/(k_i^2 k_i'{}^2)\sim
O(1)$ was used in computing the amplitude. However, our analysis
(see also \cite{Erratum}) shows that the correct estimate is:
\begin{align}\label{}
\int \frac{d^4 k}{k^2 k'{}^2}\propto \int \frac{dx\,dy\,
d^2\bk}{(mx-\bk^2+i\epsilon )(my-\bk^2+i\epsilon)}\sim
\frac{1}{m^2 E^2}\ .
\end{align}
This extra factor, for each of $n$ BHs, explains the difference between (\ref{ampl}) and the result of \cite{Vol}.

\end{document}